\documentclass[10pt,aps,twocolumn,prd,notitlepage,amssymb,amsmath,floatfix,nofootinbib,superscriptaddress]{revtex4-1}

\usepackage{epsfig}
\usepackage{bm}
\usepackage{amssymb}
\usepackage{amsmath}
\usepackage{color}
\usepackage{framed}
\usepackage{xcolor}
\usepackage[colorlinks,linkcolor=blue,anchorcolor=black,citecolor=blue]{hyperref}
\usepackage{multirow}

\usepackage[T1]{fontenc}
\usepackage[latin9]{inputenc}
\usepackage{amsmath}
\usepackage{graphicx}

\allowdisplaybreaks[4]

\mathchardef\mhyphen="2D

\makeatletter
\@ifundefined{textcolor}{}
{%
 \definecolor{BLACK}{gray}{0}
 \definecolor{WHITE}{gray}{1}
 \definecolor{RED}{rgb}{1,0,0}
 \definecolor{GREEN}{rgb}{0,1,0}
 \definecolor{BLUE}{rgb}{0,0,1}
 \definecolor{CYAN}{cmyk}{1,0,0,0}
 \definecolor{MAGENTA}{cmyk}{0,1,0,0}
 \definecolor{YELLOW}{cmyk}{0,0,1,0}
}

\makeatother
\usepackage{babel}

\begin{document}
\title{Global tensor polarization of spin $3/2$ hadrons and quark spin correlations in relativistic heavy ion collisions}

\author{Zhe Zhang}
\email{zhangzhe@mail.sdu.edu.cn}
\affiliation{Institute of Frontier and Interdisciplinary Science and Key Laboratory
of Particle Physics and Particle Irradiation (MOE), Shandong University, Qingdao, Shandong 266237, China}

\author{Ji-peng Lv}
\email{jipenglv@mail.sdu.edu.cn}
\affiliation{Institute of Frontier and Interdisciplinary Science and Key Laboratory
of Particle Physics and Particle Irradiation (MOE), Shandong University, Qingdao, Shandong 266237, China}

\author{Zi-han Yu}
\email{zihan.yu@mail.sdu.edu.cn}
\affiliation{Institute of Frontier and Interdisciplinary Science and Key Laboratory
of Particle Physics and Particle Irradiation (MOE), Shandong University, Qingdao, Shandong 266237, China}

\author{Zuo-tang Liang}
\email{liang@sdu.edu.cn}
\affiliation{Institute of Frontier and Interdisciplinary Science and Key Laboratory
of Particle Physics and Particle Irradiation (MOE), Shandong University, Qingdao, Shandong 266237, China}

\begin{abstract}
We study the global polarization of spin-$3/2$ hadrons in relativistic heavy ion collisions. 
We show in particular that the global tensor polarizations of rank two or three for spin-$3/2$ hadrons are sensitive to the local two or three quark spin correlations respectively 
in the quark gluon plasma produced in the collision processes. 
We present the relationships between these measurable tensor polarizations and quark spin correlations in the quark matter system. 
\end{abstract}
\maketitle

\section{Introduction}

The global polarization~\citep{Liang:2004ph,Liang:2004xn,Gao:2007bc} of the quark gluon plasma (QGP) in relativistic heavy ion collisions has been observed 
first in measurements of polarizations of $\varLambda$ and $\bar\varLambda$ hyperons by 
the STAR Collaboration at the Relativistic Heavy Ion Collider (RHIC)~\cite{STAR:2017ckg}  
and other experiments afterwards~\cite{STAR:2018gyt,STAR:2020xbm,STAR:2021beb,ALICE:2019onw,HADES:2022enx}.
These measurements have confirmed the theoretical predictions almost twenty years ago~\citep{Liang:2004ph,Liang:2004xn,Gao:2007bc},  
attracted new theoretical studies and lead to a new direction in Heavy Ion Physics~\citep{Betz:2007kg,Becattini:2007sr,Karpenko:2016jyx,Pang:2016igs,Li:2017slc,Xie:2017upb,Sun:2017xhx,Baznat:2017jfj,Shi:2017wpk,Xia:2018tes,Wei:2018zfb,Fu:2020oxj,Ryu:2021lnx,Fu:2021pok,Deng:2021miw,Becattini:2021iol,Wu:2022mkr,Gao:2020lxh,Huang:2020dtn,Becattini:2021lfq,Florkowski:2018fap,Becattini:2020ngo,Gao:2020vbh,Becattini:2022zvf,Jian-Hua:2023cna,Liang:2019clf,Liang:2007ma,Wang:2017jpl,Huang:2020xyr,Liang:2022ekv,Dong:2024nxj}.
Recently, the STAR Collaboration has published their measurements on the global spin alignment of vector mesons~\citep{STAR:2022fan}. 
Their data, together with other measurements~\citep{STAR:2008lcm,ALICE:2019aid,ALICE:2022byg,ALICE:2022dyy}, 
not only show that the global polarization effect manifests itself for vector mesons 
but also show that strong spin correlations exist for quarks and anti-quarks in QGP produced in the collisions. 
This leads the study on spin effects in heavy-ion collisions to a new climax~\citep{Wang:2023fvy,Chen:2023hnb,talks,Li-Juan:2023bws,Xin-Li:2023gwh,Yang:2017sdk,Sheng:2019kmk,Sheng:2020ghv,Sheng:2022wsy,Sheng:2022ffb,Xia:2020tyd,Wei:2023pdf,Kumar:2023ghs,DeMoura:2023jzz,Fu:2023qht,Sheng:2023urn,Fang:2023bbw,Dong:2023cng,Kumar:2023ojl,Lv:2024uev,Wu:2024mtj}.

A systematic formalism for describing spin correlations in a system of spin-1/2 particles 
and their relationships to experimental observables is given in~\cite{Lv:2024uev}. 
It has been pointed out that quark spin correlations in QGP can be classified as local and long range correlations in the system.  
The correlations include genuine correlations originated from the dynamical processes and induced correlations caused by averaging over other degrees of freedom in the system.
It has in particular shown that the global polarization of hyperons is sensitive to the average value of the global quark polarization, 
while the spin alignment of vector mesons is mainly determined by the local spin correlations of quark and anti-quark in the system. 

It is also interesting to see that measurements have been carried out for different vector mesons by the STAR Collaboration at RHIC~\citep{STAR:2022fan,STAR:2008lcm} 
and the ALICE Collaboration at LHC~\citep{ALICE:2019aid,ALICE:2022byg,ALICE:2022dyy}. 
The results available seem to strongly suggest that the local spin correlations between quarks and anti-quarks of different flavors are quite different from each other. 
It is thus important not only to study vector meson spin alignments but also other physical quantities to obtain information from different measurements. 
Hyperon-hyperon or hyperon-anti-hyperon spin correlations are immediately thought of such examples. 
However, it was shown in the paper~\cite{Lv:2024uev}  that they are more sensitive to the long range spin correlations. 
It would be interesting to have quantities that are sensitive to local quark-quark spin correlations. 

Recently, with developments in experiments, studies on spin-$3/2$ baryon production in high energy rections
have attracted much attention~\cite{Luk:1988as,Kim:1992az,BaBar:2008myc,Perotti:2018wxm,BESIII:2020lkm,BESIII:2023mlv,Yang:2020ezt,Zhao:2022lbw,Zhang:2023wmd,Zhang:2023box,Zhao:2024zpy,Zhang:2024rbl,Sun:2024vpp}.
Similar to vector mesons, the polarization of spin-$3/2$ baryons can be decomposed into vector and tensor polarizations.
Furthermore, in contrast to vector mesons where only rank two tensor polarizations exist, 
for spin-3/2 baryons, we have both rank two and rank three tensor polarizations. 
Such tensor polarizations can also be measured via the angular distributions of decay products. 
It is thus interesting to study these polarizations in heavy ion collisions under the quark combination mechanism. 

In this paper, we extend the study in~\cite{Lv:2024uev} to global polarizations of spin-3/2 baryons in relativistic heavy ion collisions. 
We present the relationships between the polarizations of spin-2/3 baryons and quark spin polarizations and/or correlations in QGP produced in the collisions. 
We show in particular that the tensor polarizations of spin-3/2 baryons are sensitive to the local quark-quark spin correlations in QGP 
and they provide sensitive probe to local two and three quark spin correlations in the system.  
In this connection, it is also interesting to note that a study on hypertriton polarization has been carried out at the hadronic level in~\cite{Sun:2024vpp},
where it is in particular proposed that hypertriton polarization could be used to probe its internal spin structure. 

The rest of the paper is organized as follows. 
In Sec.~\ref{sec:qc}, we briefly summarize the results of how to describe quark spin correlations in
the quark matter system proposed in~\cite{Lv:2024uev}. 
In Sec.~\ref{sec:rho3}, we present the results of the global polarization of spin-$3/2$ hadrons.
We also summarize formulae needed for measurements of these different components in Appendix A. 
Finally, a short summary and an outlook are given in Sec.~\ref{sec:summary}.

\section{quark spin correlations\label{sec:qc}}

The definitions of quark spin polarizations and correlations in QGP have been described systematically in~\cite{Lv:2024uev}. 
We summarize the main results that are used in the following of this paper in this section.

We consider a quark matter system such as QGP consisting of quarks and anti-quarks. 
The spin properties of the system are described by the spin density matrix $\hat\rho$. 
For single particle states, the spin density matrix is decomposed as 
\begin{align}
&\hat\rho^{(1)}=\frac{1}{2}(1+P_{1i}\hat\sigma_i), \label{eq:rho1/2}
\end{align} 
where $P_{1i}=\langle\hat\sigma_i\rangle={\rm Tr}[\hat\rho^{(1)}\hat{\sigma_i}]$ with 
$i=x,y,z$ is the $i$-th component of the quark polarization vector, 
$\hat\sigma_{i}$ denotes Pauli matrices and ${\rm Tr}\hat\rho^{(1)}=1$ is normalized to one.
Also, we use the convention that a sum over repeated indices is implicit through out the paper.

For two particle systems (12), 
\begin{align}
\hat\rho^{(12)}=\hat\rho^{(1)}\otimes\hat\rho^{(2)}+\frac{1}{2^2}c_{ij}^{(12)} \hat\sigma_{1i}\otimes\hat\sigma_{2j}, \label{eq:rho12}
\end{align} 
where $c_{ij}^{(12)}$ is the spin correlation between particle 1 and 2. 
Compared to the conventional decomposition in terms of 
$\{1\otimes 1, \hat\sigma_{1i}\otimes 1, 1\otimes\hat\sigma_ {2j}, \hat\sigma_{1i}\otimes\hat\sigma_{2j}\}$, 
the decomposition given by  Eq.~(\ref{eq:rho12}) has the clear advantage that $c_{ij}^{(12)}=0$ if there is no spin correlation, i.e. $\hat{\rho}^{(12)}=\hat{\rho}^{(1)}\otimes\hat{\rho}^{(2)}$. 
Here as well as in the following of this paper, we take the same scheme of notations as that in~\cite{Lv:2024uev}: 
the superscript of the spin density matrix $\hat\rho$ or the spin correlation $c$ denotes the type of particle 
where for hadrons we simply use the symbol while for quarks or anti-quarks we put it in a bracket; 
the subscript of them denotes the indices of matrix elements or spatial components.  
For polarization vectors, we simply use double subscripts to specify particle type and spatial component respectively.    

If particles in the system have other degrees of freedom that are denoted by $\alpha$,  
we consider the $\alpha$-dependence so that spin density matrices are given by 
\begin{align}
&\hat\rho^{(1)}(\alpha)=\frac{1}{2} \bigl[1+P_{1i}(\alpha) \hat\sigma_i\bigr], \label{eq:rho1/2alpha} \\
&\hat\rho^{(12)}(\alpha_1,\alpha_2)=\hat\rho^{(1)}(\alpha_1)\otimes \hat\rho^{(2)}(\alpha_2) \nonumber\\
&~~~~~~~~~~~~~~~~+\frac{1}{2^2} c_{ij}^{(12)}(\alpha_1,\alpha_2) \hat\sigma_{1i}\otimes\hat\sigma_{2j}. \label{eq:rho12alpha}
\end{align} 
For the system (12) at the state $|\alpha_{12}\rangle$, the effective spin density matrix at $\alpha_{12}$ is
\begin{align}
\hat{\bar\rho}^{(12)}&=\langle\alpha_{12}|\hat\rho^{(12)}|\alpha_{12}\rangle 
= \langle \hat\rho^{(12)} \rangle, \label{eq:rho12b}
\end{align} 
where $\langle\cdots\rangle= \sum_{\alpha_1\alpha_2} (\cdots) |\langle\alpha_{1},\alpha_{2}|\alpha_{12}\rangle|^2$ denotes the average over the state $|\alpha_{12}\rangle$. 
We decompose it in the same way  
\begin{align}
\hat{\bar\rho}^{(1)}& =\frac{1}{2} (1+\bar P_{1i} \hat\sigma_{1i} ), \label{eq:brho1} \\
\hat{\bar\rho}^{(12)}&= \hat{\bar\rho}^{(1)}\otimes \hat{\bar\rho}^{(2)}+\frac{1}{2^2}  \bar c_{ij}^{(12)}\hat\sigma_{1i}\otimes\hat\sigma_{2j} , \label{eq:rho12c} 
\end{align}
and obtain the effective polarization and correlations as
\begin{align}
&\bar P_{1i} =\langle P_{1i} \rangle, \nonumber\\
&\bar c_{ij}^{(12)} =\langle  c_{ij}^{(12)}\rangle + \bar{c}_{ij}^{(12;0)}, \label{eq:bc12} \\
&\bar{c}_{ij}^{(12;0)}= \langle P_{1i}  P_{2j}  \rangle - \bar P_{1i} \bar P_{2j} ,  \nonumber
\end{align}
where $\bar c_{ij}^{(12)}$, $c_{ij}^{(12)}$ and $\bar{c}_{ij}^{(12;0)}$ 
are called the effective, genuine and induced quark spin correlations respectively. 

Similarly, for three particle system (123) at $|\alpha_{123}\rangle$, 
\begin{align}
\hat{\bar\rho}^{(123)}=&\hat{\bar\rho}^{(1)}\otimes \hat{\bar\rho}^{(2)}\otimes\hat{\bar\rho}^{(3)} 
+\frac{1}{2^2}  \Bigl[\bar c_{ij}^{(12)}\hat\sigma_{1i}\otimes\hat\sigma_{2j}\otimes \hat{\bar\rho}^{(3)} \nonumber\\
&+\bar c_{jk}^{(23)} \hat{\bar\rho}^{(1)}\otimes\hat\sigma_{2j}\otimes\hat\sigma_{3k}  
+\bar c_{ik}^{(13)} \hat\sigma_{1i}\otimes\hat{\bar\rho}^{(2)}\otimes\hat\sigma_{3k} \Bigr] \nonumber\\
&+\frac{1}{2^3}\bar c_{ijk}^{(123)} \hat\sigma_{1i}\otimes\hat\sigma_{2j}\otimes\hat\sigma_{3k} , \label{eq:rho123b2}
\end{align} 
where the effective polarizations such as $\bar{P}_{1i}$ 
and effective two-particle spin correlations such as $\bar{c}_{ij}^{(12)}$
have similar expressions as those in the two-particle system given
by Eq.~(\ref{eq:bc12}), and the effective three-particle spin 
correlation $\bar{c}_{ijk}^{(123)}$ is given by 
\begin{align}
\bar c_{ijk}^{(123)} =& \langle  c_{ijk}^{(123)}\rangle +\bar c_{ijk}^{(123;1)} , \label{eq:bc1231} \\
\bar c_{ijk}^{(123;1)} = & \langle P_{1i}  P_{2j}  P_{3k}  + c_{ij}^{(12)} P_{3k} + c_{ik}^{(13)} P_{2j}  + c_{jk}^{(23)} P_{1i} \rangle   \nonumber\\
 - & \bar c_{ij}^{(12)} \bar P_{3k} - \bar c_{ik}^{(13)}\bar P_{2j} - \bar c_{jk}^{(23)} \bar P_{1i} 
 - \bar P_{1i} \bar P_{2j} \bar P_{3k} , \nonumber 
\end{align}
where $\bar c_{ijk}^{(123)} $ reduces to $\bar c_{ijk}^{(123;1)} $
if $c_{ijk}^{(123)}=0$ but there are two-particle spin correlations. 
It further reduces to 
\begin{align}
\bar c_{ijk}^{(123;0)}= &\langle P_{1i} P_{2j} P_{3k} \rangle -\langle P_{1i}P_{2j}\rangle \bar P_{3k} - \langle P_{1i}P_{3k}\rangle \bar P_{2j} \nonumber\\ 
&  - \langle P_{2j}P_{3k}\rangle \bar P_{1i} +2 \bar P_{1i} \bar P_{2j} \bar P_{3k} , \label{eq:bc1230} 
\end{align}
if the genuine three and two particle spin correlations all vanish. 
$\bar{c}_{ijk}^{(123;1)}$ and $\bar{c}_{ijk}^{(123;0)}$ are called the first and zeroth order induced spin correlations respectively.

\section{Polarization of spin-$3/2$ baryons}\label{sec:rho3}

The spin density matrix of spin-$3/2$ baryons is $4\times 4$ and is usually decomposed in the following way~\cite{Kim:1992az,Zhao:2022lbw} 
\begin{align}
 \hat\rho=\frac{1}{4}\left(1+\frac{4}{5} S^{i}  \hat\Sigma^{i}+\frac{2}{3} T^{ij}  \hat\Sigma^{ij}+\frac{8}{9} R^{ijk}  \hat\Sigma^{ijk}\right), \label{eq:rho2/3dec}
\end{align}
where $ \hat\Sigma^i$ is $4\times 4$ the representation of the spin operator, 
\begin{align}
 \hat\Sigma^{ij} =& \frac{1}{2}\left( \hat\Sigma^{i}  \hat\Sigma^{j}+ \hat\Sigma^{j}  \hat\Sigma^{i}\right)-\frac{5}{4} \delta^{ij} \mathbf{1},  \\
 \hat\Sigma^{ijk}=&\frac{1}{3}( \hat\Sigma^{ij}  \hat\Sigma^{k}+ \hat\Sigma^{jk}  \hat\Sigma^{i}+ \hat\Sigma^{ki}  \hat\Sigma^{j}) \nonumber\\
 &-\frac{4}{15}( \hat\Sigma^{i} \delta^{jk}+ \hat\Sigma^{j} \delta^{ik}+ \hat\Sigma^{k} \delta^{ij}). 
\end{align}
The vector polarization is $S=(0, S_{T}^{x}, S_{T}^{y},S_L)$; $T^{ij}$ is the element of the rank two tensor polarization 
and is further replaced by $S_{LL}$ and a vector $S_{LT}=(0,S_{LT}^{x},S_{LT}^{y},0)$ and a tensor with independent components 
$S_{TT}=(0,S_{TT}^{xx}, S_{TT}^{xy},0)$; 
$R^{ijk}$ is the rank three tensor polarization and the independent components are  
$S_{LLL}$; $S_{LLT}^{x}, S_{LLT}^{y}$; $S_{LTT}^{xx}, S_{LTT}^{xy}$; and $S_{TTT}^{xxx}, S_{TTT}^{xxy}$.  
A complete summary of the relationship between these independent components and elements of the spin density matrix is given in~\cite{Zhao:2022lbw}.   
These different components of polarizations can be measured in two body decays of these particles and the corresponding formulae 
are summarized in Appendix A. 

We consider the case that hadrons are produced via combination of quarks and anti-quarks in a quark matter system such as QGP. 
For the spin-3/2 baryon $B$, the production process is $q_1+q_2+q_3\to B$ and the spin density matrix is given by
\begin{align}
\hat\rho^B=\hat{\cal M}\hat\rho^{(q_1q_2q_3)}\hat{\cal M}^\dag\label{eq:3.23}
\end{align}

As discussed in~\cite{Lv:2024uev}, we start with the case where only spin degree of freedom is considered or other degrees of freedom $\alpha$ are considered 
but the wave function is factorized. 
The results obtained in both cases take the same form while in the latter case all the quantities are replaced by the effective ones. 
For explicitness, we will, as in~\cite{Lv:2024uev}, first show the results when only spin degree of freedom is considered and 
discuss the results when $\alpha$-dependence is taken into account later.  
The results obtained apply to all $J^P=(3/2)^+$ decuplet baryons. 

\subsection{With only spin degree of freedom}  \label{sec:rho3a}

In this simple case, the spin density matrix element of the baryon $B$ is given by
\begin{align}
\rho^B_{m m^{\prime}}&=\langle{j_Bm} | \hat{\mathcal{M}} \hat{\rho}^{(q_1q_2q_3)}{\hat{\mathcal{M}}}^{\dagger}| {j_Bm^{\prime}}\rangle \nonumber\\
=&\sum_{m_im^{\prime}_i} \langle{j_Bm}|\hat{\mathcal{M}} |{m_i}\rangle 
\rho^{(q_1q_2q_3)}_{m_im_i^\prime} 
\langle{m_i^{\prime}} |\hat{\mathcal{M}}^{\dagger} |{j_Bm^\prime}\rangle,  \label{eq:rhoB}
\end{align}
where $|{m_i}\rangle$ stands for $|j_1m_1,j_2m_2,j_3m_3\rangle$ for $q_1q_2q_3$; 
 $ \rho^{(q_1q_2q_3)}_{m_im_i^\prime}$ is the matrix element of the spin density matrix $\hat{\rho}^{(q_1q_2q_3)}$ of $(q_1q_2q_3)$ 
 decomposed as given by Eq.~(\ref{eq:rho123b2}). 
We emphasize that for convenience the spin quantization direction is taken as $z$ direction that corresponds to the global polarization direction. 
We use angular momentum conservation and obtain that  
\begin{align}
\langle {j_Bm}|\hat{\mathcal{M}}|{m_i}\rangle&=\sum_{j^{\prime} m^{\prime}}
\langle{j_Bm}|\hat{\mathcal{M}} | j^\prime m^\prime \rangle \langle{j^{\prime}m^{\prime}}| m_i \rangle \nonumber\\ 
&=\langle j_Bm|\hat{\mathcal{M}}|j_Bm\rangle \langle j_Bm| m_i \rangle, 
\label{eq:3.25}
\end{align}
Furthermore, the space invariance demands that $\langle j_Bm|\hat{\mathcal{M}}|j_Bm\rangle$ is independent of $m$ so that it is a constant for a given $B$ 
and can be absorbed into the normalization. 
We thus simply write,
\begin{align}
\rho^B_{m m^{\prime}}=&\sum_{m_im^{\prime}_i} 
\langle{j_Bm} |{m_i}\rangle \langle{m_i^{\prime}} |{j_Bm^\prime}\rangle  \rho^{(q_1q_2q_3)}_{m_im_i^\prime}, \label{eq:rhoB2}
\end{align}
 where $\langle{j_Bm} |{m_i}\rangle$ is the Clebsch-Gordon coefficient. 

The spin wave functions for $j_B=3/2$ hadrons are simple and the calculations are straight forward. 
We carry out these calculations and obtain the results presented in the following.  

For vector polarization, we have,
\begin{align}
S_L=&\frac{1}{2C_3} \bigl( 5\sum_{j=1}^3P_{q_jz}+ t_{zii}^{\{q_1q_2q_3\}} \bigr),       \nonumber\\
S^{x}_T=&\frac{1}{2C_3}\bigl(5\sum_{j=1}^3P_{q_jx}+ t_{xii}^{\{q_1q_2q_3\}} \bigr),    \label{eq:Sq1q2q3}   \\
S^{y}_T=&\frac{1}{2C_3}\bigl(5\sum_{j=1}^3P_{q_jy}+ t_{yii}^{\{q_1q_2q_3\}} \bigr),    \nonumber    
\end{align}
where $C_3={\rm Tr}\hat\rho^B = 3+t^{(q_1 q_2)}_{ii}+t^{(q_2q_3)}_{ii}+t^{(q_3q_1)}_{ii}$ is the normalization constant, 
and $t^{(q_1 q_2)}_{ij}$ and $t_{ijk}^{\{q_1 q_2 q_3\}}$ stand for, 
\begin{align}
t^{(q_1 q_2)}_{ij}\equiv & c^{(q_1 q_2)}_{ij}+P_{q_1i} P_{q_2j}, \\ 
t_{ijk}^{(q_1 q_2 q_3)}\equiv & c^{(q_1 q_2 q_3)}_{ijk}+c^{(q_1 q_2)}_{ij}P_{q_3k}+c^{(q_2 q_3)}_{jk}P_{q_1i}\nonumber\\ 
&+c^{(q_3 q_1)}_{ki}P_{q_2j}+P_{q_1i}P_{q_2j}P_{q_3k},\\
t_{ijk}^{\{q_1 q_2 q_3\}}\equiv & t_{ijk}^{(q_1 q_2 q_3)}+t_{ijk}^{(q_3 q_1 q_2)}+t_{ijk}^{(q_2 q_3 q_1)}. 
\end{align}
For the second rank tensor polarization, we obtain,  
 \begin{align}
   S_{LL}=&\frac{1}{C_3}\bigl[ (3t^{(q_1 q_2)}_{zz}-t^{(q_1 q_2)}_{ii})+  c(123)\bigr],  \nonumber\\ 
   S^{x}_{LT}=&\frac{3}{C_3} \bigl[ (t^{(q_1 q_2)}_{zx}+t^{(q_1 q_2)}_{xz})+  c(123)\bigr],   \nonumber\\
   S^{y}_{LT}=&\frac{3}{C_3} \bigl[ (t^{(q_1 q_2)}_{zy}+t^{(q_1 q_2)}_{yz})+  c(123)\bigr],    \label{eq:T2q1q2q3}\\
   S^{xx}_{TT}=&\frac{3}{C_3} \bigl[ (t^{(q_1 q_2)}_{xx}-t^{(q_1 q_2)}_{yy})+  c(123)\bigr], \nonumber \\
   S^{xy}_{TT}=&\frac{3}{C_3} \bigl[ (t^{(q_1 q_2)}_{xy}+t^{(q_1 q_2)}_{yx})+  c(123)\bigr].   \nonumber
   \end{align}
where $c(123)$ represents the cyclic exchange terms. 
For the third rank tensor polarization, we obtain,  
    \begin{align}
        S_{LLL}=&\frac{9}{10C_3}\bigl( 5t_{zzz}^{(q_1 q_2q_3)}  - t_{zii}^{\{q_1 q_2q_3\}}\bigr),       \nonumber\\
        S_{LLT}^{x}=&\frac{3}{10C_3}\bigl( 5t_{xzz}^{\{q_1 q_2q_3\}} - t_{xii}^{\{q_1 q_2q_3\}} \bigr),   \nonumber\\
        S_{LLT}^{y}= &\frac{3}{10C_3}\bigl( 5 t_{yzz}^{\{q_1 q_2q_3\}}  - t_{yii}^{\{q_1 q_2q_3\}}\bigr),   \nonumber\\
       S^{xx}_{LTT}=&\frac{3}{C_3} \bigl(t_{zxx}^{\{q_1q_2q_3\}}- t_{zyy}^{\{q_1q_2q_3\}} \bigr)  ,    \nonumber\\
       S^{xy}_{LTT}=&\frac{3}{C_3} \bigl(t_{zxy}^{\{q_1q_2q_3\}}+t_{zyx}^{\{q_1q_2q_3\}}\bigr)   ,        \label{eq:T3q1q2q3} \\
       S^{xxx}_{TTT}=&\frac{9}{2C_3} \bigl( t_{xxx}^{(q_1q_2q_3)}- t_{xyy}^{\{q_1q_2q_3\}} \bigr)  ,  \nonumber\\
       S^{xxy}_{TTT}=&\frac{9}{2C_3} \bigl( t_{xxy}^{\{q_1q_2q_3\}}- t_{yyy}^{(q_1q_2q_3)} \bigr)  , \nonumber
    \end{align}

Since the wave function for $J^P=(3/2)^+$ baryons are symmetric with cyclic exchange of (123), the results obtained are all in nice symmetric forms. 
They take more simplified forms if we consider the special cases for two quarks are of the same flavor. 
The physical significance can be seen more clearly so we present as example the longitudinal components in the following. 

If $q_1=q_2=q$, $q_3=q'$, we have $C_3={\rm Tr}\hat\rho^B = 3+t^{(qq)}_{ii}+2t^{(qq')}_{ii}$, 
\begin{align}
     S_L=&\frac{1}{2C_3} \bigl[ 5(2P_{qz}+P_{q'z})+ 2t_{zii}^{(qqq')} +t_{iiz}^{(qqq')} \bigr],      \nonumber \\
 %
 S_{LL}=&\frac{1}{C_3}\bigl[ (3t^{(qq)}_{zz}-t^{(qq)}_{ii})+2(3t^{(qq')}_{zz}-t^{(qq')}_{ii})  \bigr],  \label{eq:qqq'}\\ 
  %
   %
 %
 %
 %
   S_{LLL}=&\frac{9}{10C_3}\bigl(5t_{zzz}^{(qqq')} -2t_{zii}^{(qqq')}-t_{iiz}^{(qqq')} \bigr),       \nonumber 
 %
 %
\end{align}

If $q_1=q_2=q_3=q$, we have $C_3={\rm Tr}\hat\rho^B = 3(1+t^{(qq)}_{ii})$, 
\begin{align}
   S_L=&\frac{3}{2C_3} \bigl(5P_{qz}+ t_{zii}^{(qqq)} \bigr),       \nonumber\\
 %
   S_{LL}=&\frac{3}{C_3}\bigl( 3t^{(qq)}_{zz}-t^{(qq)}_{ii} \bigr),   \label{eq:qqq}\\ 
 %
 %
%
%
   S_{LLL}=&\frac{9}{10C_3}\bigl( 5t_{zzz}^{(qqq)}-3t_{zii}^{(qqq)} \bigr),        \nonumber
 %
       %
\end{align}

Suppose that we consider a system where the quark polarizations and spin correlations are small, 
and $P_{qi}$ can be considered as the first order, e.g., $\varepsilon$,  
while $c_{ij}^{(q_1q_2)}$ and $c_{ijk}^{(q_1q_2q_3)}$ are second and third order small quantities proportional to $\varepsilon^2$ and $\varepsilon^3$ respectively. 
In this case, we see that the leading contributions to the vector polarization is proportional to $P_{qi}$ with corrections proportional to $\varepsilon^3$, 
while the leading contributions to the second rank tensor polarizations are quark-quark spin correlations 
and those to the third rank tensor polarizations are three particle spin correlations of the quarks.

\subsection{With other degrees of freedom} \label{sec:rho3b}

If we have other degrees of freedom denoted by $\alpha$, we calculate all quantities at given $\alpha_B$. 
The results presented in Sec.~\ref{sec:rho3a} should be replaced by those with all quantities on the rhs being replaced by the effective ones. 
For example, $P_{qi}\to \bar P_{qi}$, $c^{(qq)}_{ij}\to \bar c^{(qq)}_{ij}$ and both the effective polarization $\bar P_{qi}$ and quark spin correlation $\bar c^{(qq)}_{ij}$ are functions of $\alpha_B$ as given by
\begin{align}
&\bar P_{qi}(\alpha_B)=\sum_{\alpha_n} |\langle \alpha_n|\alpha_B\rangle|^2 P_{qi}(\alpha_n), \label{eq:Pqeff}\\
&\bar c^{(qq)}_{ij}(\alpha_B)=\sum_{\alpha_n} |\langle \alpha_n|\alpha_B\rangle|^2 c_{ij}^{(qq)}(\alpha_n)+\bar c^{(qq;0)}_{ij}(\alpha_B). \label{eq:c2eff}
\end{align}

From the results given by Eqs~(\ref{eq:Sq1q2q3}-\ref{eq:qqq}), we see clearly that $S_L(\alpha_B)$ is sensitive to $\bar P_{qi}(\alpha_B)$, 
$S_{LL}(\alpha_B)$ is sensitive to $\bar c^{(qq)}_{ij}(\alpha_B)$ and $S_{LLL}(\alpha_B)$ is sensitive to $\bar c^{(qqq)}_{ijk}(\alpha_B)$  respectively. 
As shown by Eqs.~(\ref{eq:Pqeff}) and (\ref{eq:c2eff}), the effective quantities are results obtained after averaging over those inside the baryon $B$ in $\alpha$-space. 
Hence we see that  $S_L(\alpha_B)$ is sensitive to quark polarization averaged inside $B$, 
while $S_{LL}(\alpha_B)$ and $S_{LLL}(\alpha_B)$ are sensitive to the local two and three quark spin correlations respectively. 
Even if we further average over $\alpha_B$ in the system, we obtain the average values of the local two and three quark spin correlations respectively. 
Therefore we conclude that by tensor polarization of the second and third rank we can study local two and three quark spin correlations in the system.

The situation can be seen more clearly if we consider a simple case, i.e. if we do not distinguish between light flavors $u, d$ or $s$, 
and furthermore we take $\bar P_{qi}$ is nonzero only in the normal direction of the reaction plane as originally proposed for the global quark polarization~\citep{Liang:2004ph,Liang:2004xn,Gao:2007bc}, 
i.e. $\bar P_{qx}=\bar P_{qy}=0$, $\bar P_{qz}=\bar P_q$. 
In this simple case, we have 
\begin{align}
\bar t^{(qq)}_{ij} = & \bar c^{(qq)}_{ij}+\bar P_{q}^2\delta_{iz}\delta_{jz}, \\ 
\bar t_{ijk}^{(qqq)} = & \bar c^{(qqq)}_{ijk}+\bar P_{q}\bigl[\bar c^{(qq)}_{ij}\delta_{kz}+\bar c^{(qq)}_{ik}\delta_{jz}+\bar c^{(qq)}_{jk}\delta_{iz}\bigr] \nonumber\\
&+\bar P_q^3\delta_{iz}\delta_{jz}\delta_{kz}, \\
\bar C_3= &3(1+\bar c^{(qq)}_{ii}+\bar P_q^2),
\end{align}
and the polarizations are given by
\begin{align}
      S_L=&\frac{5}{2}\bar P_q+ \frac{3}{\bar C_3} \bigl[ \bar P_q(\bar c_{zz}^{(qq)}-2\bar c_{ii}^{(qq)}-2\bar P_q^2)+\frac{1}{2}\bar c_{zii}^{(qqq)} \bigr],       \nonumber\\
      S^{x}_T=&\frac{3}{2\bar C_3} \bigl( \bar c_{xii}^{(qqq)} +2\bar P_q\bar c_{xz}^{(qq)} \bigr),        \nonumber\\
      S^{y}_T=&\frac{3}{2\bar C_3} \bigl( \bar c_{yii}^{(qqq)} +2\bar P_q\bar c_{yz}^{(qq)} \bigr) ,       \nonumber\\
   S_{LL}=&\frac{3}{\bar C_3}\bigl( 3\bar c^{(qq)}_{zz}-\bar c^{(qq)}_{ii} +2\bar P_q^2\bigr),      \label{eq:Aqqq}\\ 
   S^{x}_{LT}=&\frac{18}{\bar C_3} \bar c^{(qq)}_{zx},    \nonumber\\
   S^{y}_{LT}=&\frac{18}{\bar C_3} \bar c^{(qq)}_{zy},    \nonumber\\
    S^{xx}_{TT}=&\frac{9}{\bar C_3} \bigl(\bar c^{(qq)}_{xx}-\bar c^{(qq)}_{yy}\bigr),  \nonumber\\ 
    S^{xy}_{TT}=&\frac{18}{\bar C_3} \bar c^{(qq)}_{xy},    \nonumber\\
%
%
   S_{LLL}=&\frac{9}{10\bar C_3}\bigl[ 5\bar c_{zzz}^{(qqq)}-3\bar c_{zii}^{(qqq)} +3\bar P_{q}(3\bar c^{(qq)}_{zz}-\bar c^{(qq)}_{ii})+2\bar P_q^3\bigr],        \nonumber\\
  S_{LLT}^{x}=&\frac{9}{10\bar C_3}\bigl(5 \bar c_{xzz}^{(qqq)}- \bar c_{xii}^{(qqq)} +8\bar P_q\bar c_{xz}^{(qq)} \bigr) ,    \nonumber\\
  S_{LLT}^{y}= &\frac{9}{10\bar C_3}\bigl(5 \bar c_{yzz}^{(qqq)} -\bar c_{yii}^{(qqq)}+8\bar P_q\bar c_{yz}^{(qq)}\bigr),    \nonumber\\
   S^{xx}_{LTT}=&\frac{9}{\bar C_3} \bigl[(\bar c_{zxx}^{(qqq)}- \bar c_{zyy}^{(qqq)}) +\bar P_q(\bar c_{xx}^{(qq)}-\bar c_{yy}^{(qq)}) \bigr]  ,     \nonumber\\
   S^{xy}_{LTT}=&\frac{18}{\bar C_3} \bigl(\bar c_{zxy}^{(qqq)}+\bar P_q\bar c_{xy}^{(qq)}\bigr),           \nonumber\\
   S^{xxx}_{TTT}=&\frac{9}{2\bar C_3} \bigl( \bar c_{xxx}^{(qqq)}- 3\bar c_{xyy}^{(qqq)} \bigr)  ,   \nonumber\\
   S^{xxy}_{TTT}=&\frac{9}{2\bar C_3} \bigl( 3\bar c_{xxy}^{(qqq)}-\bar c_{yyy}^{(qqq)} \bigr)  .    \nonumber
\end{align}

Here, we see clearly that the vector polarization is mainly determined by the quark polarization $\bar P_q$ averaged inside the baryon $B$;
the second and third rank tensor polarizations are mainly determined by the two quark spin correlation $\bar c_{ij}^{(qq)}$ 
and the three quark spin correlation $\bar c_{ijk}^{(qqq)}$ averaged inside $B$ respectively.  
We see also that the longitudinal polarization components are in general more significant than the transverse ones. 

Using the formulae presented in Appendix A, we can measure these different components of polarizations in different cases to study such local 
quark spin correlations in QGP in relativistic heavy ion collisions. 
According to the results~\cite{Lv:2024uev} of the local quark spin correlation between $s$ and $\bar s$ extracted 
from data~\cite{STAR:2022fan} on global spin alignments of $\phi$ mesons, 
spin correlations between quark and antiquark are quite strong in QGP. 
It is thus very interesting to carry out such measurements to whether there are similar spin correlations between quarks. 


\section{Summary and outlook \label{sec:summary}}

We extend the study on global polarizations and spin correlations given in~\cite{Lv:2024uev} to spin-$3/2$ hadrons 
and present the results for their global polarizations in relativistic heavy ion collisions.  
The polarizations of spin-$3/2$ hadrons are decomposed into vector, the second rank and third rank tensor polarizations. 
The results show that the vector polarizations are mainly determined by the quark polarization, 
the second rank tensor polarizations are determined by the local quark-quark spin correlations 
and the third rank tensor polarizations are determined the local spin correlations of three quarks. 

As shown in~\cite{Lv:2024uev}, when other degrees of freedom besides spin are taken into account, 
the global polarizations of produced hadrons are determined by the effective ones of quarks averaged insides the hadron. 
The quark spin correlations are mainly determined by those insides the hadron, i.e. the local spin correlations. 
Measurements of such polarization properties provide an opportunity to study local quark spin correlations 
in relativistic heavy ion collisions.
In this sense, the second and third rank tensor polarizations provide us an ideal place to study 
local two and three quark spin correlations in QGP produced in relativistic heavy ion collisions. \\[1cm]

\begin{acknowledgments}
This work was supported in part by the National Natural Science Foundation
of China under approval No. 12375075 and No.12321005, and by Shandong Province Natural Science Foundation.
\end{acknowledgments}

\begin{widetext}
\appendix
\section{Angular distributions of decay products}

For particles produced in high energy reactions, measurements of polarizations are mainly carried out 
by measuring the angular distributions of decay products in the two body decay $A\to 1+2$ in the rest frame of $A$. 
The corresponding formulae are derived using symmetry properties and conservation laws in the decay process. 
The derivations are lengthy but nevertheless straightforward. 
They can be found e.g., in~\cite{Lee:1957qs,Jacob:1959at,Rose1957,Chung:1971ri,Doncel:1972ez,Zhang:2023box}.  
For spin $3/2$ hadrons, we have not only two body decays $A\to 1+2$ but also two successive decays 
followed by $1\to 3+4$ and three successive decays followed by $3\to 5+6$. 
The angular distributions are defined as the joint distribution of that of $1$ in the the rest frame of $A$, 
and that of $3$ in the rest frame of $1$, in the case of two successive decays, and of $5$ in the rest frame of $3$ in the case of three successive decays.     
We summarize the results in the following.

(1) For $A\to 1+2$,  where spins of $A$, $1$ and $2$ are $j_A=3/2$, $j_1=1/2$, $j_2=0$, respectively, 
such as $\Delta \to N\pi$ or $\Omega\to \Lambda K$.  
We study the angular distribution of the momentum $\vec p_1$ of $1$ in the rest frame of $A$. 
Here, the coordinate system $x_Ay_Az_A$ for determining $\vec p_1$ in the rest frame of $A$ is defined in the following way: 
$z_A$ is taken in the spin quantization direction, $x_A$ and $y_A$ are defined according different conventions.
 The result for the angular distribution $W(\theta,\phi)\equiv dN/d\Omega$ is given by~\cite{Zhang:2023box} 
\begin{align}
W(\theta,\varphi)=\frac{1}{8\pi} &\Bigl\{ 2 +  (1-3\cos^2\theta)  S_{LL}
- (S_{LT}^x\cos\phi +S_{LT}^y\sin\phi ) \sin2\theta - \sin^2\theta (S_{TT}^{xx}\cos2\phi +S_{TT}^{xy}\sin2\phi ) \nonumber\\
+\alpha_A & \Bigl[ \frac{4}{5}(S_L\cos\theta+S_T^x\sin\theta\cos\phi+S_T^y\sin\theta\sin\phi) \nonumber\\
& -\frac{1}{2}(3\cos\theta+5\cos3\theta)S_{LLL}-\frac{3}{4}(\sin\theta+5\cos3\theta)(S_{LLT}^x\cos\phi+S_{LLT}^y\sin\phi) \nonumber\\
&-{3}\sin^2\theta\cos\theta (S_{LTT}^{xx}\cos2\phi+S_{LTT}^{xy}\sin2\phi) 
- \sin^3\theta (S_{TTT}^{xxx}\cos3\phi+S_{TTT}^{xxy}\sin3\phi) \Bigr]\Bigr\} . \label{eq:decH3}
\end{align}
If we integrate over $\phi$, all terms with transverse polarizations vanish and we obtain
\begin{align}
W(\theta)=\frac{1}{4} &\Bigl\{ 2 +  (1-3\cos^2\theta)  S_{LL} +\alpha_A  \Bigl[ \frac{4}{5}S_L\cos\theta  -\frac{1}{2}(3\cos\theta+5\cos3\theta)S_{LLL}  \Bigr]\Bigr\}. \label{eq:decH3b}
\end{align}
We see that we can use this to study longitudinal polarization component of $A$. 
 Alternatively, we can integrate over $\theta$ and measure the azimuthal angular distributions to study the transverse polarization components. 
 The corresponding formulae can be obtained easily and we omit it here.    
 
If parity is conserved such as $\Delta \to N\pi$, $\alpha_A=0$  and we have
\begin{align}
W(\theta,\varphi)=\frac{1}{8\pi} &\Bigl\{ 2 +  (1-3\cos^2\theta)  S_{LL}
- (S_{LT}^x\cos\phi +S_{LT}^y\sin\phi ) \sin2\theta - \sin^2\theta (S_{TT}^{xx}\cos2\phi +S_{TT}^{xy}\sin2\phi )  \Bigr\}. \label{eq:decH3c}
\end{align}
If we integrate over $\phi$ in this case, we obtain
\begin{align}
W(\theta)=\frac{1}{4} &\Bigl\{ 2 +  (1-3\cos^2\theta)  S_{LL} \Bigr\}. \label{eq:decH3d}
\end{align}
We see that it is particularly convenient to study $S_{LL}$ in this case.

(2) For two successive decays $A\to 1+2$ and $1\to 3+4$, $j_A+3/2$, $j_1=j_3=1/2$, and $j_2=j_4=0$, such as $\Omega\to\Lambda K$, $\Lambda\to p\pi^-$,
we study the joint angular distribution of $\vec p_1$ in the rest frame of $A$ and $\vec p_3$ in the rest frame of $1$. 
Here, for $\vec p_3$, the coordinate system $x_1y_1z_1$ is defined in the following way: the spin quantization direction $z_1$ of $1$ is 
taken as the helicity direction i.e. the direction of $\vec p_1$ in the rest frame of $A$;  
$y_1$ is taken as the direction of the cross product of the spin quantization direction of the mother particle $A$ with that of the produced particle $1$, 
i.e. $\hat y_1\sim \hat z_A\times \hat z_1$.   
The result for the joint distribution is quite lengthy.  
We adopt the formalism introduced in \cite{Zhang:2023box} and write
\begin{equation}
W(\theta_1,\phi_1,\theta_3,\phi_3)=\frac{1}{16 \pi^2} \sum_{n=0}^{15} w^n(\theta_1,\phi_1,\theta_3,\phi_3) S_n, \label{eq:W2}
\end{equation} 
where $S_0=1$ and $S_n (n=1-15)$ represent the 15 independent components of the spin density matrix $\hat\rho^A$ of $A$ after the decomposition.  
The results for $w_n$ are given by
\begin{align}
 w^0=& 1+\alpha_1 \alpha_A \cos{\theta_3}, \\
  w^{S_L}=&\frac{2}{5} 
    \bigl[ \cos{\theta_1}(\alpha_A+\alpha_1\cos{\theta_3})+2\alpha_1 \sin{\theta_1}\sin\theta_3 (\beta_A\sin\phi_3-\gamma_A \cos\phi_3) \bigr] ,\\
 w^{S_T^x}=&\frac{2}{5} \bigl[
    \sin\theta_1\cos\phi_1(\alpha_A+\alpha_1\cos\theta_3) -2\alpha_1 \cos\theta_1\cos\phi_1\sin\theta_3 (\beta_A \sin\phi_3 -\gamma_A  \cos\phi_3) \nonumber\\
   &~~~~~~ -2\alpha_1 \sin\phi_1 \sin\theta_3 (\beta_A\cos\phi_3 +\gamma_A \sin\phi_3) \bigr] ,\\
  w^{S_T^y} =&\frac{2}{5}\bigl[
    \sin\phi_1 \sin\theta_1(\alpha_A+\alpha_1\cos\theta_3) -2\alpha_1 \cos\theta_1 \sin\theta_3 \sin\phi_1(\beta_A \sin\phi_3 -\gamma_A \cos\phi_3) \nonumber\\
    &~~~~~~+2\alpha_1 \sin\theta_3 \cos\phi_1(\beta_A \cos\phi_3 +\gamma_A \sin\phi_3)\bigr], \\
  w^{S_{LL}}=&-\frac{1}{4} \bigl[(1+3\cos2\theta_1)(1+\alpha_1 \alpha_A \cos\theta_3)\bigr],  \\
  w^{S_{LT}^x} =& -\frac{1}{2} 
    \bigl[\sin2\theta_1 \cos\phi_1  (1+\alpha_1 \alpha_A \cos\theta_3) \bigr], \\
  w^{S_{LT}^y} =& -\frac{1}{2} 
     \bigl[\sin2\theta_1 \sin\phi_1 (1+\alpha_1 \alpha_A \cos\theta_3) \bigr], \\
   w^{S_{TT}^{xx}}=&-\frac{1}{2} 
     \bigl[\sin^2\theta_1 \cos2\phi_1 (1+\alpha_1 \alpha_A \cos\theta_3) \bigr], \\
   w^{S_{TT}^{xy}}=& -\frac{1}{2} 
     \bigl[\sin^2\theta_1 \sin2\phi_1(1+\alpha_1 \alpha_A \cos\theta_3) \bigr], \\
   w^{S_{LLL}}=&-\frac{1}{4}
    \bigl[(3\cos\theta_1+5\cos3\theta_1)(\alpha_A+\alpha_1 \cos\theta_3)
    +\alpha_1 \sin\theta_3(\sin\theta_1+5\sin3\theta_1)(\beta_A \sin\phi_3-\gamma_A \cos\phi_3)\bigr], \\
  w^{S_{LLT}^x}=&-\frac{1}{8}
    \bigl[3\cos\phi_1(\sin\theta_1+5\sin3\theta_1)(\alpha_A+\alpha_1 \cos\theta_3) 
    -\alpha_1 \sin\theta_3 \cos\phi_1(\cos\theta_1+15\cos3\theta_1)(\beta_A\sin\phi_3 -\gamma_A \cos\phi_3) \nonumber\\
    &-2\alpha_1\sin\theta_3 \sin\phi_1 (3+5\cos2\theta_1) (\beta_A \cos\phi_3 +\gamma_A \sin\phi_3)\bigr],  \\
  w^{S_{LLT}^y} =&-\frac{1}{8}
    \bigl[3\sin\phi_1(\sin\theta_1+5\sin3\theta_1)(\alpha_A+\alpha_1 \cos\theta_3)
    -\alpha_1 \sin\theta_3 \sin\phi_1(\cos\theta_1+15\cos3\theta_1)(\beta_A\sin\phi_3 -\gamma_A \cos\phi_3) \nonumber\\
    &+2\alpha_1 \sin\theta_3 \cos\phi_1 (3+5\cos2\theta_1)(\beta_A \cos\phi_3 +\gamma_A \sin\phi_3)\bigr], \\
w^{S_{LTT}^{xx}} =&-\frac{1}{8}
    \bigl[6\sin2\theta_1 \sin\theta_1 \cos2\phi_1 (\alpha_A+\alpha_1 \cos\theta_3)
    +\alpha_1 \cos2\phi_1 \sin\theta_3(\sin\theta_1-3\sin3\theta_1)(\beta_A\sin\phi_3-\gamma_A \cos\phi_3) \nonumber\\
    &-4 \alpha_1 \sin2\theta_1\sin\theta_3\sin2\phi_1 (\beta_A \cos\phi_3+\gamma_A \sin\phi_3)\bigr],    \\
w^{S_{LTT}^{xy}} =&-\frac{1}{8}
    \bigl[ 6\sin2\theta_1 \sin\theta_1 \sin2\phi_1 (\alpha_A+\alpha_1 \cos\theta_3)
   +\alpha_1 \sin2\phi_1 \sin\theta_3(\sin\theta_1-3\sin3\theta_1) (\beta_A\sin\phi_3-\gamma_A \cos\phi_3) \nonumber\\
    &+4 \alpha_1 \sin2\theta_1\sin\theta_3\cos2\phi_1(\beta_A \cos\phi_3+\gamma_A\sin\phi_3)\bigr], \\
  w^{S_{TTT}^{xxx}} =& -\frac{1}{2}\sin^2\theta_1
    \Bigl\{\sin\theta_1 \cos3\phi_1(\alpha_A+\alpha_1 \cos\theta_3)  \nonumber\\
    &-\alpha_1\sin\theta_3 \bigl[ \cos\theta_1 \cos3\phi_1  (\beta_A \sin\phi_3-\gamma_A \cos\phi_3) 
    +\sin3\phi_1 (\beta_A \cos\phi_3+\gamma_A \sin\phi_3)\bigr] \Bigr\}, \\
 w^{S_{TTT}^{xxy}} =&-\frac{1}{2} \sin^2\theta_1
    \Bigl\{ \sin\theta_1 \sin3\phi_1 (\alpha_A+\alpha_1\cos\theta_3)  \nonumber\\
    &-\alpha_1 \sin\theta_3 \bigl[ \cos\theta_1\sin3\phi_1 (\beta_A \sin\phi_3-\gamma_A \cos\phi_3) 
       - \cos3\phi_1 (\beta_A \cos\phi_3+\gamma_A \sin\phi_3)\bigr] \Bigr\}. 
    \label{eq:4.112}
\end{align}
If we integrate over $\phi_1$ and $\phi_3$, again all terms with transverse polarization vanish and we obtain
\begin{align}
W(\theta_1,\theta_3)=&\frac{1}{4} 
\Bigl\{ (1+\alpha_1 \alpha_A \cos{\theta_3} ) \bigl[ 1- \frac{1}{4} S_{LL} (1+3\cos2\theta_1)\bigr]
+\frac{2}{5} S_L \cos{\theta_1}(\alpha_A+\alpha_1\cos{\theta_3})  \nonumber\\
&-\frac{1}{4}S_{LLL} (3\cos\theta_1 +5\cos3\theta_1)(\alpha_A +\alpha_1\cos\theta_3) \Bigr\}. \label{eq:W2b}
\end{align}
If we further integrate over $\theta_3$, Eq.~(\ref{eq:W2b}) just reduces to Eq.~(\ref{eq:decH3b}).

(3) For three successive decays $A\to 1+2$, $1\to 3+4$ and $3\to 5+6$,  $j_A=3/2$, $j_1=j_3=j_5=1/2$, and $j_2=j_4=j_6=0$, we consider only
$\Xi^*\to\Xi\pi$, $\Xi\to\Lambda\pi$, and $\Lambda\to p\pi$, where $\Xi^*\to\Xi\pi$ is a parity conserved strong decay.
Here, for $\vec p_5$, the coordinate system $x_3y_3z_3$ is defined in the similar way as $x_1y_1z_1$: 
the spin quantization direction $z_3$ of $3$ is taken as the helicity direction i.e. the direction of $\vec p_3$ in the rest frame of $1$;  
$y_3$ is taken as the direction of the cross product of the spin quantization direction of the mother particle $1$ with that of the produced particle $3$, 
i.e. $\hat y_3\sim\hat z_1\times \hat z_3$.   
We take again the formalism introduced in \cite{Zhang:2023box} and write the distribution in the same form as that given by Eq.~(\ref{eq:W2}). 
The coefficients are given by the following equations.  
\begin{align}
  w^0=&1+\alpha_{\Lambda} \alpha_{\Xi} \cos{\theta_p}, \\
  w^{S_L}=&\frac{2}{5}\Bigl\{(\alpha_{\Xi}+\alpha_{\Lambda} \cos{\theta_p})
    (\cos{\theta_{\Xi}} \cos{\theta_{\Lambda}} -2\sin{\theta_{\Lambda}}
    \sin{\theta_{\Xi}}\cos{\phi_{\Lambda}}) \nonumber \\
    &+\alpha_{\Lambda} \sin{\theta_p} \Bigl[ (2\cos{\theta_{\Lambda}}
    \cos{\phi_{\Lambda}}\sin{\theta_{\Xi}}+\cos{\theta_{\Xi}} \sin{\theta_{\Lambda}}) (\beta_{\Xi} \sin{\phi_p}-\gamma_{\Xi} \cos{\phi_p}) 
    \nonumber\\
    &~~~+2 \sin{\theta_{\Xi}} \sin{\phi_{\Lambda}}(\beta_{\Xi} \cos{\phi_p}+\gamma_{\Xi} \sin{\phi_p}) \Bigr] \Bigr\}, 
    \\
   w^{S^x_T}=&\frac{2}{5}\Bigl\{ (\cos{\theta_{\Lambda}} \cos{\phi}_{\Xi} \sin{\theta_{\Xi}}
        +2\cos{\theta_{\Xi}} \cos{\phi_{\Xi}} \cos{\phi_{\Lambda}} \sin{\theta_{\Lambda}}-2\sin{\theta_{\Lambda}}
        \sin{\phi_{\Xi}} \sin{\phi_{\Lambda}})(\alpha_{\Xi}+\alpha_{\Lambda} \cos{\theta_p}) 
        \nonumber\\
        &- \alpha_{\Lambda} \sin{\theta_p} \Bigl[2(\sin{\phi_{\Xi}}
        \cos{\phi_{\Lambda}}+\cos{\theta_{\Xi}} \cos{\phi_{\Xi}} \sin{\phi_p})
        (\beta_{\Xi} \cos{\phi_p} +\gamma_{\Xi} \cos{\phi_p}) 
        \nonumber\\
        &~~~ - (\sin{\theta_{\Xi}} \cos{\phi_{\Xi}} \sin{\theta_{\Lambda}}
        +2 \sin{\phi}_{\Xi} \cos{\theta_{\Lambda}} \sin{\phi_{\Lambda}}
        -2\cos{\theta_{\Xi}} \cos{\phi_{\Xi}} \cos{\theta_{\Lambda}} \cos{\phi_{\Lambda}}) 
        \nonumber\\ 
        &~~~\times(\beta_{\Xi}\sin{\phi_p}-\gamma_{\Xi} \cos{\phi_p})\Bigr]\Bigr\}, 
        \\
      w^{S^y_T}=&\frac{2}{5}\Bigl\{ (\cos{\theta_{\Lambda}} \sin{\phi}_{\Xi} \sin{\theta_{\Xi}}
        +2\cos{\theta_{\Xi}} \sin{\phi_{\Xi}} \cos{\phi_{\Lambda}} \sin{\theta_{\Lambda}}+2\sin{\theta_{\Lambda}}
        \cos{\phi_{\Xi}} \sin{\phi_{\Lambda}})(\alpha_{\Xi}+\alpha_{\Lambda} \cos{\theta_p})
        \nonumber\\
        &+ \alpha_{\Lambda} \sin{\theta_p}\Bigl[ 2(\cos{\phi_{\Xi}}\cos{\phi_{\Lambda}}-\cos{\theta_{\Xi}} \sin{\phi_{\Xi}} \sin{\phi_p})
        (\beta_{\Xi} \cos{\phi_p} +\gamma_{\Xi} \cos{\phi_p})
        \nonumber\\
        &~~~+(\sin{\theta_{\Xi}} \sin{\phi_{\Xi}} \sin{\theta_{\Lambda}}
        -2 \cos{\phi}_{\Xi} \cos{\theta_{\Lambda}} \sin{\phi_{\Lambda}}-2\cos{\theta_{\Xi}} \sin{\phi_{\Xi}} \cos{\theta_{\Lambda}} \cos{\phi_{\Lambda}})
        \nonumber\\
        &~~~\times (\beta_{\Xi} \sin{\phi_p}-\gamma_{\Xi} \cos{\phi_p})\Bigr]\Bigr\},   
         \\
          w^{S_{LL}}=&-\frac{1}{4}(1+3\cos2{\theta_{\Xi}})(1+\alpha_{\Lambda} \alpha_{\Xi} \cos{\theta_p}),\\
           w^{S^x_{LT}}=&-\frac{1}{2}\sin{2\theta_{\Xi}} \cos{\phi}_{\Xi} (1+\alpha_{\Lambda} \alpha_{\Xi} \cos{\theta_p}), 
          \\
          w^{S^y_{LT}}=&-\frac{1}{2}\sin{2\theta_{\Xi}} \sin{\phi_{\Xi}} (1+\alpha_{\Lambda} \alpha_{\Xi} \cos{\theta_p}),
          \\
          w^{S^{xx}_{TT}}=&-\frac{1}{2}\sin^2{\theta_{\Xi}}\cos{2\phi_{\Xi}} (1+\alpha_{\Lambda} \alpha_{\Xi} \cos{\theta_p}), 
          \\
          w^{S^{xy}_{TT}}=&-\frac{1}{2}\sin^2{\theta_{\Xi}}\sin{2\phi_{\Xi}} (1+\alpha_{\Lambda} \alpha_{\Xi} \cos{\theta_p}), 
          \\
          w^{S_{LLL}}=&-\frac{1}{4} \Bigl\{ (\alpha_{\Xi} +\alpha_{\Lambda} \cos{\theta_p}) 
          \bigl[\cos{\theta_{\Lambda} } (3\cos{\theta_{\Xi}}+5\cos{3\theta_{\Xi}}) -\cos{\phi_{\Lambda}} \sin{\theta_{\Lambda}} (\sin{\theta_{\Xi}}+5\sin{3\theta_{\Xi}})  \bigr]
          \nonumber\\
    &+\alpha_{\Lambda} \sin{\theta_p}
    \Bigl[ \bigl[\cos{\theta_{\Lambda}} \cos{\phi_{\Lambda}}(\sin{\theta_{\Xi}}
     +5\sin{3\theta_{\Xi}})+(3\cos{\theta_{\Xi}}+5\cos{3\theta_{\Xi}}) \sin{\theta_{\Lambda}}\bigr]  (\beta_{\Xi} \sin{\phi_p}-\gamma_{\Xi} \cos{\phi_p})  
    \nonumber\\
    &+ (\sin{\theta_{\Xi}}+5\sin{3\theta_{\Xi}}) \sin{\phi_{\Lambda}}
    (\beta_{\Xi} \cos{\phi_p}+\gamma_{\Xi} \sin{\phi_p}) \Bigr] \Bigr\},
    \\
    w^{S^x_{LLT}}=&\frac{1}{8} \Big \{ (\alpha_{\Xi}+\alpha_{\Lambda} \cos{\theta_p})
    \bigl[-3\cos{\theta_{\Lambda}} \cos{\phi_{\Xi}}(\sin{\theta}_{\Xi}+5\sin{3\theta_{\Xi}})
    -\sin{\theta_{\Lambda}}(\cos{\theta_{\Xi}}+15\cos{3 \theta_{\Xi}}) \cos{\phi_{\Xi}} \cos{\phi}_{\Lambda}  
    \nonumber\\
    &~~~+2 \sin{\theta_{\Lambda}} (3+5\cos2{\theta_{\Xi}}) \sin{\phi_{\Xi}} \sin{\phi_{\Lambda}} \bigr]
    \nonumber\\
    &+\alpha_{\Lambda} \sin{\theta_p} \Bigl[ \bigl[2(3+5\cos{2\theta_{\Xi}}) \sin{\phi_{\Xi}}
    \cos{\phi_{\Lambda}} +(\cos{\theta_{\Xi}}+15 \cos{3 \theta_{\Xi}})\cos{\phi_{\Xi}} \cos{\phi_{\Lambda}}\bigr] (\beta_{\Xi} \cos{\phi_p}+\gamma_{\Xi} \sin{\phi_p})
    \nonumber\\
    &~~~+\bigl[ (\cos{\theta_{\Xi}}+15 \cos{3 \theta_{\Xi}}) \cos{\phi_{\Xi}} \cos{\theta_{\Lambda}} \cos{\phi_{\Lambda}}
      -2(3+5 \cos{2\theta_{\Xi}}) \sin{\phi_{\Xi}} \cos{\theta_{\Lambda}} \sin{\phi_{\Lambda}} 
    \nonumber\\
   &~~~-3(\sin{\theta_{\Xi}}+5\sin{3\theta_{\Xi}}) \cos{\phi_{\Xi}} \sin{\theta_{\Lambda}} \bigr] (\beta_{\Xi} \sin{\phi_p} -\gamma_{\Xi} \cos{\phi}_P) \Bigr]\Big \} ,
    \\
 w^{S^y_{LLT}}=&\frac{1}{8} \Big \{ (\alpha_{\Xi}+\alpha_{\Lambda} \cos{\theta_p})
    \bigl[ -3\cos{\theta_{\Lambda}} \sin{\phi_{\Xi}}(\sin{\theta}_{\Xi}+5\sin{3\theta_{\Xi}})
    -\sin{\theta_{\Lambda}}(\cos{\theta_{\Xi}}+15\cos{3 \theta_{\Xi}}) \sin{\phi_{\Xi}} \sin{\phi}_{\Lambda}      \nonumber\\
    &~~~ -2 \sin{\theta_{\Lambda}} (3+5\cos2{\theta_{\Xi}}) \cos{\phi_{\Xi}} \sin{\phi_{\Lambda}} \bigr]      \nonumber\\
    &-\alpha_{\Lambda} \sin{\theta_p} \Bigl[ \bigl[ 2(3+5\cos{2\theta_{\Xi}}) \cos{\phi_{\Xi}}
    \cos{\phi_{\Lambda}} -(\cos{\theta_{\Xi}}+15 \cos{3 \theta_{\Xi}})\sin{\phi_{\Xi}} \sin{\phi_{\Lambda}}]
    (\beta_{\Xi} \cos{\phi_p}+\gamma_{\Xi} \sin{\phi_p})      \nonumber\\ 
    &~~~+ \bigl[ 2(3+5 \cos{2\theta_{\Xi}}) \cos{\phi_{\Xi}} \cos{\theta_{\Lambda}} \sin{\phi_{\Lambda}}
    -3(\sin{\theta_{\Xi}}+5\sin{3\theta_{\Xi}}) \sin{\phi_{\Xi}} \sin{\theta_{\Lambda}}     \nonumber\\
    &~~~+(\cos{\theta_{\Xi}}+15 \cos{3 \theta_{\Xi}}) \sin{\phi_{\Xi}} \cos{\theta_{\Lambda}} \cos{\phi_{\Lambda}} \bigr]
     (\beta_{\Xi} \sin{\phi_p} -\gamma_{\Xi} \cos{\phi}_P) \Bigr] \Big \} , 
     \\
     w^{S^{xx}_{LTT}}=&-\frac{1}{4} \sin{\theta_{\Xi}} \Big \{ (\alpha_{\Xi}
     +\alpha_{\Lambda} \cos{\theta_p}) \Bigl[\cos{2\phi_{\Xi}}\bigl[ 3\cos{\theta_{\Lambda}} \sin{2\theta_{\Xi}} 
     + (1+3\cos{2\theta_{\Xi}}) \cos{\phi_{\Lambda}} \sin{\theta_{\Lambda}} \bigr]       \nonumber\\
     &~~~-4\cos{\theta_{\Xi}} \sin{\theta_{\Lambda}} \sin{2\phi_{\Xi} }  \sin{\phi_{\Lambda}} \Bigr]        \nonumber\\
     &-\alpha_{\Lambda} \sin{\theta_p}\Bigl[ \bigl[4 \sin{2 \phi_{\Xi}} \cos{\theta_{\Xi}} \cos{\phi_{\Lambda}}
     +(1+3 \cos{2 \theta_{\Xi}}) \cos{2\phi_{\Xi}} \sin{\phi_{\Lambda}}\bigr] (\beta_{\Xi} \cos{\phi_p}+\gamma_{\Xi} \sin{\phi_p})
     \nonumber\\
     &~~~ - \bigl[3 \sin{2\theta_{\Xi}} \cos{2\phi_{\Xi}} \sin{\theta_{\Lambda}}+4 \cos{\theta_{\Xi}} \sin{2\phi_{\Xi}} \cos{\theta_{\Lambda}} \sin{\phi_{\Lambda}}
     \nonumber\\
     &~~~ - (1+3 \cos{2 \theta_{\Xi}}) \cos{2\phi_{\Xi}} \cos{\theta_{\Lambda}} \cos{\phi_{\Lambda}}\bigr] 
     (\beta_{\Xi} \sin{\phi_p}-\gamma_{\Xi} \cos{\phi_p}) \Bigr] \Big \}, 
     \\
     w^{S^{xy}_{LTT}}=&-\frac{1}{4} \sin{\theta_{\Xi}} \Big \{(\alpha_{\Xi}+\alpha_{\Lambda} \cos{\theta_p})
     \bigl[\sin{2\phi_{\Xi}}(3\cos{\theta_{\Lambda}} \sin{2\theta_{\Xi}}+ (1+3\cos{2\theta_{\Xi}}) \cos{\phi_{\Lambda}}  \sin{\theta_{\Lambda}} ) 
     \nonumber\\
     &~~~+4\cos{\theta_{\Xi}} \sin{\theta_{\Lambda}} \cos{2\phi_{\Xi} }  \sin{\phi_{\Lambda}}\bigr] 
     \nonumber\\
     &+\alpha_{\Lambda} \sin{\theta_p} \Bigl[
     \bigl[4 \cos{2 \phi_{\Xi}} \cos{\theta_{\Xi}} \cos{\phi_{\Lambda}}-(1+3 \cos{2 \theta_{\Xi}}) \sin{2\phi_{\Xi}} \sin{\phi_{\Lambda}} \bigr]
     (\beta_{\Xi} \cos{\phi_p}+\gamma_{\Xi} \sin{\phi_p})
     \nonumber\\
     &~~~+ \bigl[3 \sin{2\theta_{\Xi}} \sin{2\phi_{\Xi}} \sin{\theta_{\Lambda}}-4 \cos{\theta_{\Xi}} \cos{2\phi_{\Xi}} \cos{\theta_{\Lambda}} \sin{\phi_{\Lambda}}
     \nonumber\\
     &~~~ -(1+3 \cos{2 \theta_{\Xi}}) \sin{2\phi_{\Xi}} \cos{\theta_{\Lambda}} \cos{\phi_{\Lambda}} \bigr]
     (\beta_{\Xi} \sin{\phi_p}-\gamma_{\Xi} \cos{\phi_p}) \Bigr] \Big \} , 
     \\
     w^{S^{xxx}_{TTT}} =& \frac{1}{2} \sin^2{\theta_{\Xi}}
       \Big \{     (\alpha_{\Xi}+\alpha_{\Lambda} \cos{\theta_p})
       \bigl[\sin{3  \phi_{\Xi}} \sin{\phi_{\Lambda}} \sin{\theta_{\Lambda}}
       -\cos{3 \phi_{\Xi}} (\cos{\theta_{\Lambda}} \sin{\theta_{\Xi}}+\cos{\theta_{\Xi}} \cos{\phi_{\Lambda}} \sin{\theta_{\Lambda}})\bigr] 
       \nonumber\\
     &+\alpha_{\Lambda} \sin{\theta_p} \Bigl[
       (\sin{3 \phi_{\Xi}} \cos{\phi_{\Lambda}}
       +\cos{3 \phi_{\Xi}} \cos{\theta_{\Xi}} \sin{\phi_{\Lambda}})(\beta_{\Xi} \cos{\phi_p}+\gamma_{\Xi} \sin{\phi_p})
       \nonumber\\
       &~~~ - \bigl[\cos{3 \phi_{\Xi}} (\sin{\theta_{\Xi}} \sin{\theta_{\Lambda}}- \cos{\theta_{\Xi}} \cos{\theta_{\Lambda}} \cos{\phi_{\Lambda}})
       +\sin{3\phi_{\Xi}} \cos{\theta_{\Lambda}} \sin{\phi_{\Lambda}}\bigr]
       (\beta_{\Xi} \sin{\phi_p} -\gamma_{\Xi} \cos{\phi_p})  \Bigr] \Big \}, 
       \\
       w^{S^{xxy}_{TTT}} =& -\frac{1}{2} \sin^2{\theta_{\Xi}}
       \Big \{  (\alpha_{\Xi}+\alpha_{\Lambda} \cos{\theta_p})
      \bigl[ \cos{3  \phi_{\Xi}} \sin{\phi_{\Lambda}} \sin{\theta_{\Lambda}}
       +\sin{3 \phi_{\Xi}} (\cos{\theta_{\Lambda}} \sin{\theta_{\Xi}}+\cos{\theta_{\Xi}} \cos{\phi_{\Lambda}} \sin{\theta_{\Lambda}}) \bigr]  
       \nonumber\\
       &+\alpha_{\Lambda} \sin{\theta_p} \Bigl[
       -(\sin{3 \phi_{\Xi}} \cos{\theta_{\Xi}} \sin{\phi_{\Lambda}}-\cos{3 \phi_{\Xi}} \cos{\phi_{\Lambda}}) 
       (\beta_{\Xi} \cos{\phi_p}+\gamma_{\Xi} \sin{\phi_p})
       \nonumber\\
       &~~~+\bigl[\sin{3\phi_{\Xi}}(\sin{\theta_{\Xi}} \sin{\theta_{\Lambda}}-\cos{\theta_{\Xi}} \cos{\theta_{\Lambda}} \cos{\phi_{\Lambda}})
       -\cos{3\phi_{\Xi}} \cos{\theta_{\Lambda}} \sin{\phi_{\Lambda}}\bigr]
       (\beta_{\Xi} \sin{\phi_p} -\gamma_{\Xi} \cos{\phi_p}) \Bigr]
       \Big \} .
 \end{align}

Again, if we integrate over the azimuthal angles $\phi_\Xi, \phi_\Lambda$ and $\phi_p$, all transverse polarization dependent terms vanish. 
We obtain 
\begin{align}
W(\theta_\Xi,\theta_\Lambda,\theta_p)=&\frac{1}{8} 
\Bigl\{1+\alpha_{\Lambda} \alpha_{\Xi} \cos{\theta_p}
+\frac{2}{5}S_L (\alpha_{\Xi}+\alpha_{\Lambda} \cos{\theta_p}) \cos{\theta_{\Xi}} \cos{\theta_{\Lambda}}  \nonumber\\
&-\frac{1}{4} S_{LL} (1+3\cos2{\theta_{\Xi}})(1+\alpha_{\Lambda} \alpha_{\Xi} \cos{\theta_p}) \nonumber\\
 &-\frac{1}{4} S_{LLL} (\alpha_{\Xi} +\alpha_{\Lambda} \cos{\theta_p})
    \cos{\theta_{\Lambda} } (3\cos{\theta_{\Xi}}+5\cos{3\theta_{\Xi}}) \Bigr\}. 
     \label{eq:W3b}
\end{align}
It is also clear that if we integrate over $\theta_p$, Eq.~(\ref{eq:W3b}) reduces to Eq.~(\ref{eq:W2b}) with $\alpha_A=0$. 

\end{widetext}



\end{document}